\documentclass[aps,prl,twocolumn,showpacs,amsmath,amssymb,superscriptaddress]{revtex4-1}

\usepackage{graphicx}
\usepackage{color}
\usepackage{listings}
\usepackage{braket}

\begin{document}
\title{Topological monopoles and currents in electromagnetic waves}

\author{Vladlen Shvedov$^1$, Wieslaw Krolikowski}
\affiliation{Laser Physics Centre, Research School of
Physics and Engineering, Australian National University, Canberra, ACT
0200, Australia}
\affiliation{Science Program, Texas A\&M University at Qatar, Doha, Qatar}

\begin{abstract}
 Singularities, i.e. places of discontinuities of parameters are extremely general objects appearing in electromagnetic waves and thus are the key to understanding fundamental wave processes. These structures commonly occur in purely coherent, highly directional waves, such as laser beams, determining additional spatial or "`topological"' properties of the wave fields independently of their propagational dynamics. For instance, topologies of wave fronts, called phase singularities, add orbital degrees of freedom to electromagnetic waves. These singularities are common to all types of scalar waves described only by their intensity and phase distributions. As the electromagnetic wave is a vector wave, its topological properties generally depend on all field components leading to complex field patterns in space and time. These patterns may contain singular points of undefined instantaneous orientation of the vector field, i.e.  – the instantaneous field (IF) singularities. In zero-order paraxial approximation of purely  transverse electromagnetic waves, some instantaneous field distributions may carry apparent topological monopoles, when the originally source-free wave exhibits spatial structure associated with the 2D "virtual" sources of electromagnetic fields.

Here we present systematic description of topological singularities in both electric and magnetic field of the electromagnetic waves in a paraxial approximation. We also consider the important types of paraxial electromagnetic waves with complex transverse field structures containing instantaneous field singularities.

\end{abstract}
\maketitle

\section{introduction}
The traditional approach to describe propagating electromagnetic radiation in free space is based on the differential form of Maxwell's electromagnetic theory and the corresponding wave equation [1]. This approach gives a detailed account of electromagnetic waves as synchronized transverse oscillations of electric and magnetic fields satisfying the superposition principle at a point. As result, all fundamental wave parameters, such as amplitude, 
wave frequency, direction in space (wave vector), orientation of fields (polarization) and phase refer to behavior of the wave at a particular point in space and time. This representation of electromagnetic wave via only local "at point" parameters is perfect from the fundamental point of view on condition that the wave field is free of physical discontinuities.
There is no doubt that the local wave parameters can be different at different points of the wave [2-10] even in purely coherent and highly directional electromagnetic radiations, such as paraxial laser beams [11-13]. It is also true that as the state of the wave field changes, the parameters may not always be defined throughout space. Typically, there are some particular points in the wave field where the local parameters, such as the phase, state and handedness of the polarization or orientation of electric (magnetic) vectors are not defined [4]. These points are known as singularities in the wave field [14]. 
As all certain values of a singular parameter in the wave occur in the neighbourhood of a singular point (but not in the point itself), singularities define spatial properties of the surrounding wave field [9]. 
This requires one to deal with additional ``integral'' characteristics describing spatial``topological'' properties of the wave field [4, 5]. They are related to the fundamental geometrical aspects of electromagnetic waves and are associated with the inhomogeneous distributions of local parameters, such as the phase and electric (magnetic) vector orientation, which constitute the wave singularities. 
These singularities  are typically classified as phase and polarization singularities, depending on which parameter is undefined (or singular) at some points in the wave. Phase singularities, commonly known in light fields as ``optical vortices" [15], are features of scalar components of vectorial electromagnetic waves, leading to a circulation of the electromagnetic energy around the point in two dimensions [16-19]. They also occur naturally for all types of scalar waves, both classical and quantum [20, 21], and have been studied in detail in a number of works in the different branches of wave physics [see, for instance vortices in electron beams [22-24].

Other types of singularities may occur in electromagnetic waves when the vectorial properties of the waves cannot be ignored. Well known are the polarization singularities that naturally appear in completely polarized electromagnetic waves where the state of polarization varies with position, first described in detail by J. Nye [25, 26]. They refer to the points were the direction of the polarization or the handedness of the polarization ellipse is not defined [9, 25]. As the state of polarization is actually determined by  electromagnetic vectors averaged over the period of wave oscillations, the polarization pattern alone cannot properly expose the deep nature of singularities in these fields. The polarization provides only rough and, sometimes, distorted picture of instantaneous orientation of electric vectors in the vicinity of a singular point in the vector wave. The simplest example is a vector field of a uniformly circularly (generally elliptically) polarized optical vortex which cannot be properly explained in terms of either pure phase or polarization singularities, as we will show in Section 2. 
On the other hand, the instantaneous orientations of electromagnetic vectors in the wave filed provide a natural and efficient way to describe the most fundamental features of physical discontinuities in paraxial vector waves. Corresponding instantaneous vector field (IF) singularities  are not stable, but exhibit periodic evolution in propagation in space as the field vectors oscillate in time. These special IF singularities are source of unique 2D field structures which we call \textit{wave monopoles} and \textit{wave currents} in transverse components of the wave. We will consider them in detail in Section 3. Then, in Section 4, we will discuss wave fields with mixed topology and polarization singularities.
To clarify the concept of IF singularities, we will start from the general principles of electromagnetic wave topology and its trivial cases of homogeneously (generally elliptically) polarized fields and then we will move on to more complex scenarios involving scalar and vector singularities in the waves.  Moreover, we consider here only electromagnetic waves which carry finite energy flux and are self-similar in space. The former leads to spatially  localized waves in the form of light beams. The latter means that the intensity of the beam is similar in each  plane $\textit{S}_\bot$ orthogonal to the propagation direction.

\section{Paraxial model of electromagnetic beams}

In general, the electromagnetic fields {\textbf{E}, \textbf{B}} in a free propagating wave are complicated functions of both space and time. The situation becomes simpler for monochromatic harmonic waves with  constant  frequency $\omega$, were one can naturally separate spatial and time dependences. In many practical cases, this allows one to describe a harmonic wave as a product of complex-valued spatial mode and  a complex exponential of time, remembering to take the real part to obtain the physical fields:  $\{{\bf E,B}\}$=Re$\{{\bf F}(x,y,z)e^{-i\omega t}\}$.
Description  of spatial modes is based on the spatially invariant Helmholtz wave equation (which in turn is derived from differential form of Maxwell  equations)
 \begin{equation}\label{GrindEQ__1_}
\left(\nabla^2+k^2\right){\bf F}=0,
\end{equation}
where  $\nabla={{\bf e}_ x}\partial_x + {{\bf e}_y}\partial_y+ {{\bf e}_z}\partial_z,$ $\partial_u\equiv\partial/\partial u$, \textit{u}=\textit{x},\textit{y},\textit{z}, and $ {\bf e}_x, {\bf e}_y, {\bf e}_z$, are the unit vectors in Cartesian coordinates, $k = \omega/c$  is a wave number, c is speed of light, and {\bf F} represents  electric or magnetic field, respectively. 
When the electromagnetic radiation is a highly directional along a certain spatial axis (for example, \textit{z}), a spatial mode can be expressed as  ${\bf F}(x,y,z)=\widetilde{\bf F}(x,y,z)e^{ikz}$ with a complex amplitude   $\widetilde{\bf F}(x,y,z)$ satisfying to the spatially asymmetric equation
 \begin{equation}\label{GrindEQ__3_}
\left(\nabla^2+2ik\partial_z\right)\widetilde{\bf F}=0.
\end{equation}

As was shown by M. Lax {\em et al.} [27], in the zero-order paraxial approximation, the electromagnetic field are purely transverse $\bf{\widetilde{F}}=\bf{F}_\bot$  and the second \textit{z}-derivative could be ignored in the Eq.(2) with respect to a slowly-varying  first \textit{z}-derivative of the amplitude function $\textbf{F}_\bot$. Finally, the equation for the complex amplitude takes  form of a parabolic equation:
	\begin{equation}
\left(\nabla^2_\bot+2ik\partial_z\right){\bf F}_\bot=0,
\end{equation}
where $\nabla_\bot={{\bf e}_ x}\partial_x + {{\bf e}_y}\partial_y$.

Note that the transversality of the field breaks down in the next order correction  [27]. By analysing the higher-order terms in the series expansion of Maxwell's equations, Lax found that the first-order terms of expansion has a small longitudinal field component and its value is obtained from the transverse components through the relation $\textit{k}\textit{F}_\textit{z}=\textit{i}\nabla_\bot\textbf{F}_\bot$. The role of longitudinal components in the next order of the paraxial approximation will be discussed in sections 4 and 5. Here, our main interest is the zero-order paraxial approximation as the basis to clarify  the topology and field singularities phenomena in the coherent electromagnetic waves.
A scalar form of the Eq. (3) is represented by a parabolic equation of the Schr\"{o}dinger type that describes the complex amplitude of a Cartesian component $\textit{F}_\bot=\textit{F}_{\textit{x},\textit{y}}$ of the electromagnetic wave
	\begin{equation}
\nabla^2_\bot{\textit{F}}_\bot=-2ik\partial_z{\textit{F}}_\bot.
\end{equation}
 						
This equation admits solutions in a form of self-similar beams preserving their structure in propagation [28]. Many of such solutions with both, infinite and finite energies, have been obtained in the last decades [13, 29-36]. Main attention has been paid to the spatially-localized solutions with finite energy flux [29-35] as the only ones having physical sense.  They include, among others,   the well-known Hermite-Gaussian and Laguerre-Gaussian beams; elegant and generalized Laguerre-Gaussian beams; Bessel-Gauss beams; hypergeometric-Gauss beams; fractional-order elegant Laguerre-Gaussian beams; Bessel-Gauss beams with quadratic radial dependence; Mathieu and Ince-Gaussian beams, and optical vortex beams (see [36] and references therein).

As indicated by  their names, all spatially localized characteristics of such beams are predominantly associated with the Gaussian envelope
\begin{equation} \label{GrindEQ__5_} 
G(x,y,z)=\frac{1}{\xi (z)} \exp \left(-\frac{\rho ^{2} }{w_{0}^{2} \xi (z)} \right),                                                 
\end{equation} 
\noindent where $\rho =\sqrt{x^{2} +y^{2} } $ is the radial coordinate of an arbitrary point, $\xi (z)=1+{iz\mathord{\left/ {\vphantom {iz z_{0} }} \right. \kern-\nulldelimiterspace} z_{0} } $, $z_{0} ={kw_{0}^{2} \mathord{\left/ {\vphantom {kw_{0}^{2}  2}} \right. \kern-\nulldelimiterspace} 2} $ and \textit{w}${}_{0}$ is a beam waist in the plane \textit{z~}=~0.

The scalar function \textit{G} is extremely important in understanding paraxial electromagnetic wave behavior because it serves as a background envelope for more general wave solutions, ensuring finiteness of their energy as $\int _{0}^{\infty }(2/\pi w_{0}^{2} )GG^{*} dS_{\bot }=1$ (the star denotes complex conjugation).

The general, rigorous solutions to the parabolic equation \eqref{GrindEQ__3_} can be represented as a product [38]

\begin{equation} \label{GrindEQ__6_} 
\textbf{F}_{\bot } =\textbf{U}(x,y,z)G(x,y,z),
\end{equation} 
 where \textbf{U}(\textit{x,y,z})\textbf{ }is a complex vector modulating function and G(\textit{x,y,z}) is of the form of Eq.\eqref{GrindEQ__5_}.
While the Gaussian amplitude describes evolution of the localized envelope of scalar waves or beams in space, the modulation function \textbf{U} determines their all topological and vector properties. Despite the infinite number of possible modulation functions \textbf{U},\textbf{\textit{ }} their general structure can be defined by just a few basic topological properties. Therefore, in the following we will concentrate on the functions {\textbf U} having the most typical topological properties.  However, before going any further, let us clarify the relevant terminology. 

As the functions \textbf{U} are vector fields defining the orientation of electric (magnetic) fields in space, they are directly related to the state of polarization of the beams. Typically, only beams with spatially variant polarization in their transverse cross-sections are deemed vector beams. In contrast, beams with a constant polarization are generally considered as scalar.

At this point, it is worth clarifying the relationship between the state of polarization and field orientation as these two terms that are often, yet incorrectly, used interchangeably. The state of polarization is a time-averaged parameter of a beam and, so far, can be purely homogeneous while the instantaneous orientation of electric (magnetic) vectors provides a complex pattern in the transverse cross-section of the beam. In such cases, the electromagnetic fields of the beam are solutions to the full vector wave equation, and cannot be defined by one scalar function. According to superposition principle, it is always possible to present the transverse vector function \textbf{U} (corresponding to \textbf{E} or \textbf{B} fields) in terms of their Cartesian components: 

\begin{equation} \label{GrindEQ__7_} 
\textbf{U}=A_{x} U_{x} (x,y,z)\textbf{e}_{x} +A_{y} U_{y} (x,y,z)\textbf{e}_{y},     
\end{equation} 
where  \textit{A${}_{x}$} and \textit{A${}_{y}$} are arbitrary, generally complex constants.

We consider the beam to be a ``scalar'' when  its electromagnetic field can be described by the real
part of a single scalar modulating function:
%
\begin{equation} \label{GrindEQ__8_} 
\frac{Re\left\{A_{x} U_{x} \right\}}{Re\left\{A_{x} \right\}} =\frac{Re\left\{A_{y} U_{y} \right\}}{Re\left\{A_{y} \right\}}=Re\left\{U\right\},
\end{equation}
In case of a ``vector beam'' the transverse Cartesian components of the modulating function \eqref{GrindEQ__7_} \textbf{U}\textit{${}_{x}$}=\textit{A${}_{x}$U${}_{x}$}(\textit{x},\textit{y},\textit{z})
\textbf{e}\textit{${}_{x}$, }and\textit{ }\textbf{U}\textit{${}_{y}$}=\textit{A${}_{y}$U${}_{y}$}(\textit{x},\textit{y},\textit{z})\textbf{e}\textit{${}_{y}$} depend in different way on coordinates:
\begin{equation} \label{GrindEQ__9_} 
\frac{Re\left\{A_{x} U_{x} \right\}}{Re\left\{A_{x} \right\}} \ne \frac{Re\left\{A_{y} U_{y} \right\}}{Re\left\{A_{y} \right\}}, 
\end{equation} 

The simplest example of the scalar beam is an elliptically polarized Gaussian beam. This trivial case occurs if the modulating function is constant: $\textbf{U}=\textit{A}_x\textbf{e}_x+\textit{A}_y\textbf{e}_y$, where $\textit{A}_x=\textit{a}_x$ and $\textit{A}_y=\textit{a}_y e^{i\phi}$ ; $\textit{a}_x$ and $\textit{a}_y$ are real numbers and $\phi$ is the phase difference between Cartesian components of \textbf{U}. Then one immediately obtains the  simplest solution of Eq.\eqref{GrindEQ__3_} in  form of complex amplitude $\textbf{F}_\bot=(\textit{a}_x\textbf{e}_x+\textit{a}_y e^{i\phi}\textbf{e}_y)\textit{G}$ of the elliptically polarized beam $\textbf{ F}_\bot\textit{e}^{i(kz-\omega t)}$,  which certainly is a scalar wave. This paraxial, fundamental Gaussian beam [11-13], is only one of the infinite numbers of possible variations of electromagnetic beams described by Eq. \eqref{GrindEQ__6_}. However, homogeneous distributions of instantaneous electric (magnetic) field vectors with the same orientation at each point of the transverse plane of beams is a sufficiently rare. Generally, beams have complex distributions of phase, polarization or the instant directions of the electromagnetic fields changing  from point to point in the transverse plane. 

In the next Section, we will provide examples of  scalar and vector beams with nontrivial topology of modulating functions.  

\section{Scalar and vector beams with topological charges}
\noindent In general, the function $\textbf{U}(\rho,\varphi,z)$ depends on all three coordinates. By changing variables as $\textit{R}_{t} =\sqrt{X^{2} +Y^{2} } =\rho/(\textit{w}_0\xi)$, where $\textit{X}=\textit{x}\slash(\textit{w}_0\xi);\textit{Y}=\textit{y}\slash(\textit{w}_0\xi)$, following  standard procedure  shown in works [34, 38], we insert expression \eqref{GrindEQ__6_} into wave equation \eqref{GrindEQ__3_} and obtain differential equation for the modulating function $\bf{U}(\textit{R}_\textit{t},\varphi,\xi)$ in the form
\begin{equation} \label{GrindEQ__10_} 
\nabla _{t}^{2} \textbf{U}-4\xi ^{2} \partial _{\xi } \textbf{U}=0 
\end{equation} 
where $\nabla_\textit{t}=\textbf{e}_\textit{x}\mathrm{\partial}_\textit{X}+\textbf{e}_\textit{y}\mathrm{\partial }_\textit{Y}$.

\noindent This equation admits the separation of variables

\begin{equation} \label{GrindEQ__11_} 
\textbf{U}(R_{t} ,\varphi ,\xi )=\textbf{u}(R_{t} ,\varphi )Z(\xi ),       
\end{equation} 

\noindent where the function $\textit{Z}(\xi)=\exp(\textit{K}^{2}(1-\xi)\mathrm{\slash }4\xi)$, and \textit{K} is  a constant. Then the function $\bf{u}(\textit{R}_\textit{t},\varphi)$ satisfies  the two dimensional vector Helmholtz equation

\noindent 
\begin{equation} \label{GrindEQ__12_} 
\left(\nabla _{t}^{2} +K^{2} \right) \textbf{u}=0.
\end{equation} 

In order to analyze possible solutions of Eq. \eqref{GrindEQ__12_} let us start with an important particular case when the constant \textit{K} is equal to zero and, hence \textit{Z~}=1. Then the modulation function \textbf{U} is independent of the longitudinal coordinate $\xi$ and \textbf{U}=\textbf{u}. In this case, the vector Helmholtz equation \eqref{GrindEQ__12_} reduces to the  vectorial Laplace equation for the function $\bf{U}(\textit{R}_\textit{t},\varphi)$ 

\begin{equation} \label{GrindEQ__13_} 
\nabla _{t}^{2} \textbf{U}(R_{t} ,\varphi )=0.
\end{equation}

\noindent Components \textit{U${}_{x}$} and \textit{U${}_{y}$} of the modulating function \textbf{U} are, in general, independent scalar functions, interpreted as complex amplitudes of electromagnetic waves 
along Cartesian coordinates \textit{x} and \textit{y}, respectively.  When the polarization state of the light field is homogenous (generally elliptical), the modulating function \textbf{U} can be represented by a single scalar functions \textit{U${}_{x}$}(\textit{x},\textit{y},\textit{z})=\textit{U${}_{y}$}(\textit{x},\textit{y},\textit{z})= \textit{U}(\textit{x},\textit{y},\textit{z}) as

\noindent 
\begin{equation} \label{GrindEQ__14_} 
\textbf{U}=\left(a_{x} \textbf{e}_{x} +a_{y} e^{i\phi } \textbf{e}_{y} \right) U,      
\end{equation} 
\noindent where \textit{U} satisfies to the two-dimensional Laplace equation

\noindent 
\begin{equation} \label{GrindEQ__15_}
\nabla _{t}^{2} U=0 
\end{equation} 

\begin{figure}[htbp]
\centering
\includegraphics[width=8.0cm]{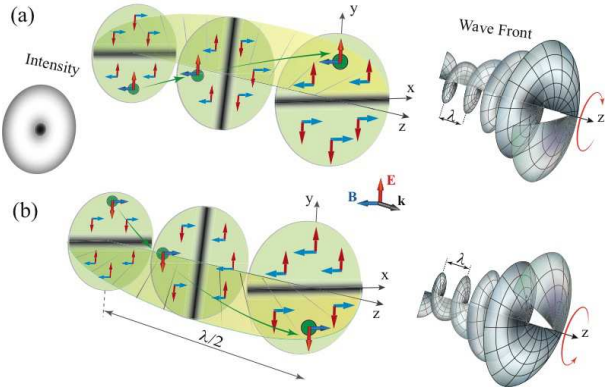}
\caption{\label{Fig1} \@ The instantaneous, near-axis distribution of the electromagnetic fields and wave front of the scalar, linearly polarized vortex with topological charge $ l=+1$ (a)  and $ l=-1$~(b). Each point of the wave's surface (see green circle) follows a spiral trajectory around the axis of the beam as the latter propagates. }
\end{figure}

\noindent From the spectrum of possible solutions of Eq. \eqref{GrindEQ__15_}, well known as harmonic functions [39], we will consider the simplest but critically important solutions which play a key role in classification of light singularities in scalar light beams

\noindent 
\begin{equation} \label{GrindEQ__16_} 
U(R_{t} ,\varphi )=R_{t}^{n} \exp (isn\varphi ),
\end{equation}

\noindent where \textit{n~}=~0;$\pm$1;$\pm$2{\dots}, and  \textit{s}=$\pm$1. 

\noindent The function \eqref{GrindEQ__16_} has singularity on the line \textit{R${}_{t}$}=0 when \textit{n~}$\mathrm{\neq }$~0, and its modulus on this line is either 0 for \textit{n~}$>$ 0 or $\mathrm{\infty }$ for \textit{n} $<$~0. Finite energy solution \eqref{GrindEQ__6_} with modulation function \eqref{GrindEQ__16_} can be realized for\textit{ n~}$\mathrm{\ge }$~0, except for some specific superpositions in which the self-similarity of the resulting light beam is lost [40]. Then the product \textit{sn} in an exponent of Eq.\eqref{GrindEQ__16_} can be represented by  a new integer\textit{ l~}=~0;$\pm$1;$\pm$2{\dots}, and order \textit{n} in \textit{R${}_{t}$} becomes\textit{ n~}=~{\textbar}\textit{l}{\textbar}, when \textit{n~}$\mathrm{\ge }$0. The physical solutions now read

\noindent 
\begin{equation} \label{GrindEQ__17_} 
U(R_{t} ,\varphi )=R_{t}^{\left|l\right|} \exp (il\varphi ).
\end{equation}

\noindent While the harmonic function \eqref{GrindEQ__17_} is the simplest solution of Eq. \eqref{GrindEQ__15_} it gives full information about the topological structure in the field's components. The azimuthal index \textit{l} defines the topological structure of the wave surface. In optics this index is named a topological charge of an optical vortex. The term ``optical vortex'' refers to the general solution \eqref{GrindEQ__6_} with scalar modulation functions containing addend $\textit{l}\varphi$ in its phase, $\theta(\rho,\varphi,z)$. Then the wave front of the optical vortex has a form of a helicoid-like surface twisted around the singular line $\textit{R}_{t}=0$. Moreover, the phase circulating around the singularity line in the function \eqref{GrindEQ__17_} results in phase discontinuity that is multiple of the topological charge

\noindent 
\begin{equation} \label{GrindEQ__18_} 
\oint d\theta  =\oint \nabla \theta  d\textbf{l}=2\pi l ,
\end{equation} 
where \textit{d}\textbf{l} is the infinitesimal vector element of the loop. Consequently, the directional  electromagnetic wave with the modulating function \eqref{GrindEQ__17_} carries  discrete angular momentum [5] along the propagation axis \textit{z}

\begin{equation} \label{GrindEQ__19_} 
M_{z} ={\left\langle U^{*} \left|L_{z} \right|U\right\rangle \mathord{\left/ {\vphantom {\left\langle U^{*} \left|L_{z} \right|U\right\rangle  \left\langle U^{*} |U\right\rangle }} \right. \kern-\nulldelimiterspace} \left\langle U^{*} |U\right\rangle } =l\hbar , 
\end{equation} 
where $\it{L_z}={-}\it{i}\hbar\partial/\partial\varphi$ is the generator of rotations around \textit{z}. 

\noindent The above discussed features  apply to   any  scalar modulating functions with the singular phase factor $\textit{l}\varphi$. 

In general, the scalar Cartesian components of the function $\textbf{u}(\textit{R}_\textit{t},\varphi)$ in Eq. \eqref{GrindEQ__11_} satisfy the two-dimensional Helmholtz equation \eqref{GrindEQ__12_} which takes a scalar form
\begin{equation} \label{GrindEQ__20_} 
\left(\nabla _{t}^{2} +K^{2} \right) u=0 .
\end{equation} 
Among an infinite number of possible solutions of Eq. \eqref{GrindEQ__20_} there are axially symmetric solutions with the phase topology similar to \eqref{GrindEQ__17_}

\begin{equation} \label{GrindEQ__21_} 
u(R_{t} ,\varphi )=J_{|l|} (KR_{t} )\exp (il\varphi ) 
\end{equation} 
where  \textit{J${}_{l}$~}(\textit{x}) is the\textit{l}-order Bessel function of the first kind.
The corresponding scalar components of modulating function \eqref{GrindEQ__11_} are

\begin{equation} \label{GrindEQ__22_} 
U(R_{t} ,\varphi ,\xi )=J_{\left|l\right|} (KR_{t} )Z(\xi )\exp (il\varphi ) .
\end{equation}

\noindent These functions constitute basis of scalar Bessel-Gauss beams with a phase singularity of a type of optical vortex located along the optical axis (when \textit{l~}$\mathrm{\neq}$~0). 


\noindent According to Eqs. \eqref{GrindEQ__6_} and \eqref{GrindEQ__14_}, the electric fields of arbitrarily homogeneously polarized Gaussian and Bessel-Gauss types of vortex beams with modulating functions Eqs. \eqref{GrindEQ__17_} and \eqref{GrindEQ__22_} are [41]

\noindent 
\begin{equation} \label{GrindEQ__23_} 
\begin{array}{l} {\textbf{E}_{\bot } =\text{Re}\left\{\left(E_{0x} \textbf{e}_{x} +E_{0y} e^{i\phi } \textbf{e}_{y} \right)\right. \; } \\ 
{\; \; \; \; \; \times \left. R\; _{t}^{\left|l\right|} e^{il\varphi } G(\rho ,z)e^{i(kz-\omega t)} \right\}}, \end{array} 
\end{equation} 
\begin{equation} \label{GrindEQ__24_} 
\begin{array}{l} {\textbf{E}_{\bot }^{J} =\text{Re}\left\{\left(E_{0x} \textbf{e}_{x} +E_{0y} e^{i\phi } \textbf{e}_{y} \right)\right. \; } \\ {\times \left. J_{\left|l\right|} (KR_{t} )Z(\xi )e^{il\varphi } G(\rho ,z)e^{i(kz-\omega t)} \right\}}, \end{array} 
\end{equation}

\noindent where \textit{E}${}_{0}$\textit{${}_{x}$} and \textit{E}${}_{0}$\textit{${}_{y}$} are real constant amplitudes of the \textit{x-} and \textit{y}-components of the electric field, respectively. By introducing a new constant $\beta=\textit{k}\,{\sin}\alpha$ in Eq. \eqref{GrindEQ__22_} as the angular parameter of the wave vector \textbf{k}, one can define $\textit{K}=\beta\textit{w}_0$, and $Z(\xi )=\exp \{ -i\beta ^{2} z/(2k\xi )\} $. When \textit{l}=0 the Eq. \eqref{GrindEQ__24_} represents the well-known 
zero order Bessel-Gauss beam [37, 42].
The  magnetic field component corresponding to field (\ref{GrindEQ__24_}) can be uniquely determined  by $c\textbf{B}_{\bot } =\textbf{e}_{z} \times \textbf{E}_{\bot } $.

Following our earlier discussion (see Eqs. 8 and 9), in general, the two-component fields \eqref{GrindEQ__23_} and \eqref{GrindEQ__24_} are vector fields because of different phase factors, $\cos(\textit{l}\varphi)$ and $\sin(\textit{l}\varphi)$, in their Cartesian components Re$\{$\textit{U${}_{x}$}$\}$$\mathrm{\neq }$ Re$\{$\textit{U${}_{y}$}$\}$. Exception are the beams with zero topological charge \textit{l}=0 and the beams with purely homogeneous linear states of polarization (when $\phi=\textit{m}\pi$, $\textit{m}=0;1;2...)$. They include elliptically polarized fundamental Gaussian and Bessel-Gauss beams (described in the end of the previous section) which are both typical scalar beams.  The latter are scalar optical vortex beams as their electric field oscillates in a single direction and they can be described by real part of a single scalar function.

Figures 1 (a) and (b) show examples of the electromagnetic field distribution with function \eqref{GrindEQ__23_} corresponding to the linearly (\textit{y}-polarized) Gaussian beams with topological charges \textit{l }= 1 and \textit{l~}=$\mathrm{-}$1, respectively. The  wave fronts of these scalar beams  continuously twist such that  the field pattern rotates around the beam axis. The pattern contains the zero-field twisted surface which separates the out-of-phase fields. In general, the number of such surfaces corresponds to the topological charge carried by the beam.

As we mentioned above, the fields \eqref{GrindEQ__23_} and \eqref{GrindEQ__24_} represent, in general, elliptically polarized vector beams. Their modulating vector functions \eqref{GrindEQ__14_} admits separation of variables
\begin{equation} \label{GrindEQ__25_} 
\textbf{U}(R_{t} ,\varphi ,\xi )=f(R_{t} ,\xi )\textbf{g}(\varphi ) ,
\end{equation} 
with function $\textbf{g}(\varphi)$ 
\begin{equation} \label{GrindEQ__26_} 
\textbf{g}(\varphi )=\left(a_{x} \textbf{e}_{x} +a_{y} e^{i\phi } \textbf{e}_{y} \right)e^{i(l\varphi +\varphi _{0} )} ,     
\end{equation} 

\noindent and $\textit{f}=\textit{R}^{|\textit{l}|}$ for the Laplace's type, and $\textit{f}=\textit{J}_{|\textit{l}|}(\textit{K}R_t)\textit{Z}(\xi)$ for the Helmholtz type of solutions \eqref{GrindEQ__23_}, respectively. 
Let us look in detail at the particular but interesting case of circularly polarized vortex beams which can be obtained directly from the vector function \eqref{GrindEQ__26_} for $\textit{a}_x=\textit{a}_y=\sqrt2$ and $\phi=\pm\pi/2$. Then the function \eqref{GrindEQ__26_} becomes
\begin{equation} \label{GrindEQ__27_} 
\textbf{g}(\varphi )=e^{i(l\varphi +\varphi _{0} )} \pmb{\sigma} ,
\end{equation} 
where $\pmb{\sigma}=(\textbf{e}_\textit{x} +\textit{i}\sigma\textbf{e}_\textit{y})/\sqrt2$ is  a unit vector and $\sigma=\pm1$ with positive (negative) sign  corresponding  to right (left) hand polarization,  and $\varphi_0$ is an arbitral initial phase.

\noindent In the more explicit form, the equation \eqref{GrindEQ__27_} may be represented in the polar basis of radial and azimuthal unit vectors $\textbf{e}_\rho$ and $\textbf{e}_\varphi$ as:

\begin{equation} \label{GrindEQ__28_} 
\textbf{g}(\varphi )=(\textbf{e}_{\rho } +i\sigma \; \textbf{e}_{\varphi } )\; \exp \{ i\; [(l+\sigma )\varphi +\varphi _{0} ]\} .
\end{equation} 

\noindent Distributions of electric and magnetic fields in vector vortex beams corresponding to the modulating function \eqref{GrindEQ__28_} with $\sigma\textit{l}=1$ and $\sigma\textit{l}=-1$ are shown in Figure 2. One can see that to in contrast to scalar case, these vector fields are spatially inhomogeneous and evolve over the period of the fields oscillations.  As this spatial evolution depends on both topological charge \textit{l} and handedness $\sigma$, the vector vortex beams carry topology- and polarization-dependent discrete angular momentum along their propagation [43]

\noindent
\begin{equation} \label{GrindEQ__29_} 
M_{z} =(l+\sigma )\hbar  .
\end{equation} 

It follows from Eq. \eqref{GrindEQ__29_} that the field rotation around the phase singularity can be decreased or increased by varying circulation of the field vectors at a point. 

\begin{figure}[htbp]
\centering
\includegraphics[width=7.0cm]{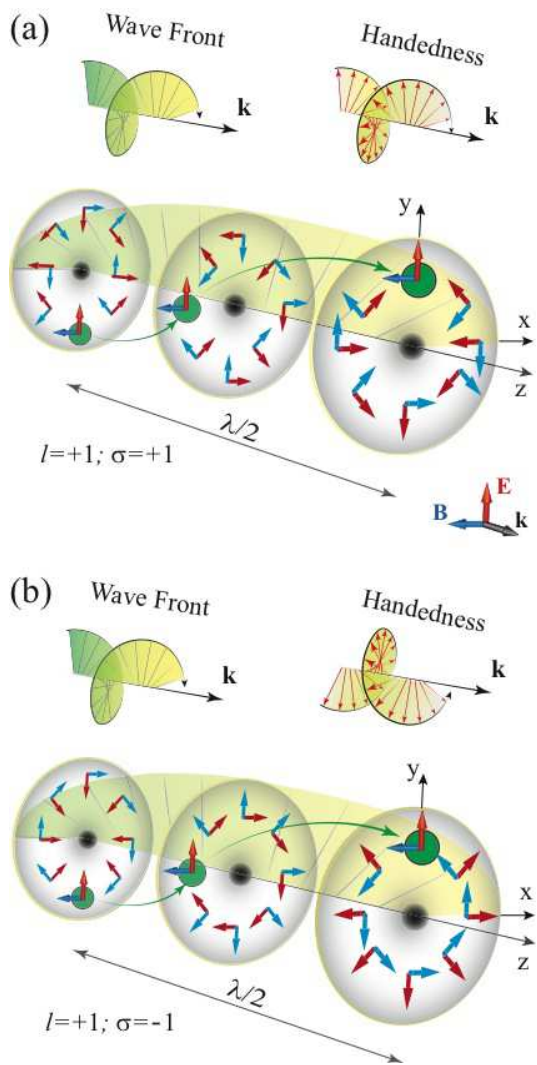}
\caption{\label{Fig2}\@    The instantaneous near-axis distribution of the electromagnetic field in the vector circularly polarized optical vortex with topological charge and handedness a) l =+1, $\sigma $ =+1; and b) l =+1, $\sigma$ =-1. The evolution of a fixed arbitrary point on the wave surface is shown with green circle.}
\end{figure}

To conclude this paragraph, let us reiterate the most relevant points. As it clearly seen in Figures 1 and 2 the instantaneous distribution of electromagnetic fields is dramatically different in scalar and vector vortex beams. The field pattern of scalar beams consists of lines of discontinuities in transverse plane at each instant  of time. These lines rotate with wave's frequency leading to the time average null intensity along the axis (line of singularity). In contrast, the field pattern of vector beam always possesses a point singularity (null of the  field) in its cross-section. This type of discontinuity, which we call a vector singularity, will be discussed in detail in the next Section. As for now, let us note that modulating functions of both vector and scalar vortex beams may not be rigidly coupled in space with the envelope Gaussian wave \eqref{GrindEQ__5_} which serves as a background for topological charges. For clarity we have chosen the specific case of topological charges and the singularity lines located along at the maximum of the Gaussian function \eqref{GrindEQ__5_} (along the beam axis). Generally, the wave field can have many separated topological charges with sufficiently arbitrary trajectories. The singularity lines may extend over three dimensions, and be embedded in the volume filled by the background Gaussian function either forming closed loops or infinite, unbounded lines [44-47].

\noindent The singularities essentially change spatial topology of a wave field and, as consequence, the overall intensity pattern. They are the cause of local energy redistributions in the wave beam [19]. However, as the singularities exist only in the host Gaussian envelope \eqref{GrindEQ__5_}, they do not influence the general hyperbolic character of the beam self-diffraction. The topological modulating function itself has a physical sense only in combination with the background function which is responsible for the finite energy flux through the cross section of the beam. In other words, the function \eqref{GrindEQ__5_} generates "space" for existence of topological objects in a wave field and it ensures that the spatially unlimited paraxial beam carries a finite power.

\section{Vector topologies in paraxial electromagnetic waves}

When a vortex beam is elliptically polarized, the competition occurs between the phase topology in scalar components of its electromagnetic field and the effect of field orientations due to the wave polarization, as the latter already cannot be ignored [48]. This leads to rich topological patterns of the instantaneous electromagnetic field. The field rotates at every fixed point in space (except the point of singularity) with the wave frequency, while the twisted wave front of the beam rotates with this frequency around a point of the phase singularity (see Fig.2). The result of the both local (at point) and total (in space) rotations is an inhomogeneous distribution of the electromagnetic field in any transverse plane of the vortex beam.  Despite the homogeneous state of polarization, the instantaneous field pattern of the beam contains vector singularities as points of an undefined orientation of the fields. 
In this section, we will discuss these vectorial topological objects  which we named instantaneous field (IF) singularities.

Unlike the elliptically polarized vortex beam, the time-averaged pattern of an inhomogeneously  distributed field may  lead to a nontrivial polarization distribution  which  may vary in space exhibiting  the so-called polarization singularities. These  singularities have been   considered extensively in the literature [9,25, 26, 49-63]. 
Here, we will  focus   on a specific  situation when a beam  is linearly polarized at any point, but the polarization direction  varies azimuthally. In such beams  the local rotation of the electric (magnetic) field at a point cancel out  the rotation of total field caused by the phase singularity. This mutual cancellation  can be achieved through a superposition of waves with opposite topological charges and orthogonal states of polarizations. 
 The idea to use superposition of four uniformly linearly polarized vortex beams to  create nonuniformly polarized waves   was first proposed by F. Gori  [41].  It was further developed, with little variations,  to construct more general  complex polarization structures  [62, 64-70]. Following this technique, we consider  the  circularly polarized  field with modulating functions Eqs. \eqref{GrindEQ__27_}, \eqref{GrindEQ__28_}.  Let us introduce the superposition of these fields having opposite signs of topological charge \textit{l } and handedness $\sigma$, as

\begin{equation} \label{GrindEQ__30_} 
\textbf{g}(\varphi )=\pmb{\sigma} ^{\pm } e^{i(l\varphi +\varphi _{0} )} +\pmb{\sigma} ^{\mp } e^{-i(l\varphi +\varphi _{0} )} .
\end{equation} 
The corresponding electric and magnetic fields of Gaussian beams are
 
\begin{equation} \label{GrindEQ__31_} 
\begin{array}{l} {\textbf{E}_{\bot } =E_{0} \text{Re}\left\{R_{t}^{\left|l\right|} G(\rho ,z)e^{i(kz-\omega t)} \right\}} \\ {\; \; \; \; \; \times \left(\cos \left[(l+\sigma )\varphi +\varphi _{0} \right]\textbf{e}_{\rho } \right. } \\ {\; \; \; \; \left. -\sigma \sin \left[(l+\sigma )\varphi +\varphi _{0} \right]\textbf{e}_{\varphi } \right)} \\\\ 
{\textbf{B}_{\bot } =B_{0} \text{Re}\left\{R_{t}^{\left|l\right|} G(\rho ,z)e^{i(kz-\omega t)} \right\}\ } \\ {\; \; \; \; \; \times \left(\sigma \sin \left[(l+\sigma )\varphi +\varphi _{0} \right]\textbf{e}_{\rho } \right. } \\ {\; \; \; \; \; +\left. \cos \left[(l+\sigma )\varphi +\varphi _{0} \right]\textbf{e}_{\varphi } \right)} ,\end{array} 
\end{equation}
 where \textit{B}${}_{0}$ =\textit{E}${}_{0}$$\mathrm{\slash }$\textit{c}.

\noindent The  field of vector Bessel-Gauss beams can be found in the similar way

\begin{equation} \label{GrindEQ__32_} 
\begin{array}{l} {\textbf{E}_{\bot }^{J} =E_{0} \text{Re}\left\{J_{\left|l\right|} (KR_{t} )Z(\xi )G(\rho ,z)e^{i(kz-\omega t)} \right\}} \\ {\; \; \; \; \; \times \left(\cos \left[(l+\sigma )\varphi +\varphi _{0} \right]\textbf{e}_{\rho } \right. } \\
 {\; \; \; \; -\left. \sigma \sin \left[(l+\sigma )\varphi +\varphi _{0} \right]\textbf{e}_{\varphi } \right)} \\\\ 
 {\textbf{B}_{\bot }^{J} =B_{0} \text{Re}\left\{J_{\left|l\right|} (KR_{t} )Z(\xi )G(\rho ,z)e^{i(kz-\omega t)} \right\} } \\ {\; \; \; \; \; \times \left(\sigma \sin \left[(l+\sigma )\varphi +\varphi _{0} \right]\textbf{e}_{\rho } \right. } \\ {\; \; \; \; \; +\left. \cos \left[(l+\sigma )\varphi +\varphi _{0} \right]\textbf{e}_{\varphi } \right)} .\end{array} 
\end{equation} 

\begin{figure}[htbp]
\centering
\includegraphics[width=8.0cm]{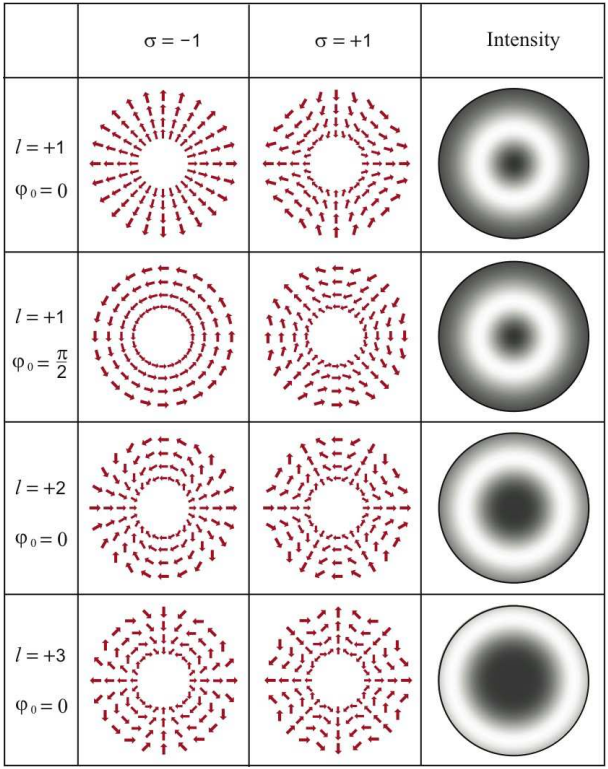}
\caption{\label{Fig3} \@  Examples of electric field distribution (second and third columns) and intensity (fourth column) of vector beams described by Eq. (31).  
Only the intensity of  x-polarized component of the beams is shown. }
\end{figure}

\noindent Both Eq.\eqref{GrindEQ__31_} and Eq.\eqref{GrindEQ__32_} types of beams possess similar topological properties. Hence, without loss of generality, we will concentrate below on the Gaussian beams expressed by Eq.(31).  Figure 3 depicts few examples of electric field distribution in such beams. These patterns correspond to pure linear states of polarization at any point of a transverse plane, but the direction of the polarization depends on the azimuthal position. The angular and spin momenta of the beams are zero and, consequently,  the numbers \textit{l} and $\sigma$ in Eqs.(31, 32) can no longer be interpreted as topological charge and spin, respectively. Nevertheless, these numbers have a new meaning as indices within the symmetry group in the beams. The group with the same index \textit{l} splits into two subgroups  according to  $\sigma$. These subgroups \textit{l}$\sigma$ have own symmetry with unique distribution of the electric (magnetic) field and cannot be transformed  from one to another without symmetry breaking. Each subgroup \textit{l}$\sigma$$>$0 or \textit{l}$\sigma$$<$0 with a fixed \textit{l} permits local rotations of fields by the same angle at every point of the wave front by changing initial phase $\varphi_0$. The local rotations by  the angle $\varphi_0$ are equal to a total rotation of the electric (magnetic) field pattern by  the angle $\varphi_0$/2 around the beam center preserving field symmetry in every subgroup except \textit{l}$\sigma$={-}1 (see Fig. 3).

\noindent Despite the absence of topological charges in the vector beams Eqs.\eqref{GrindEQ__31_},\eqref{GrindEQ__32_}, the field configuration still carries  IF singularities originating from the undefined direction of the \textit{instantaneous} field vectors. They may have a source, spiral, sink, saddle or circulation morphology as  shown in Figure 3. In particular case described in this Section, the IF singularities are isolated points at the axis $\rho$=0 in a beam transverse plane. Their position and time-average morphology coincide with vector point singularities (\textit{V}-points) proposed by I. Freund as points at which the orientation the electric (magnetic) vectors of a linearly polarized field becomes undefined [52]. However, there are significant differences between them. The \textit{V}-points are singularities in the state of polarization rather than fields and, thus, as it was embraced by I. Freund, they always belong to linearly polarized waves. In contrast, the IF singularities may exist in any state of polarization. The example is the singularities in circularly polarized optical vortices shown in Section 2 (see Fig. 2 and the text below). Their electric (magnetic) fields contain IF singularities at every moment of time (see Fig. 3). Another difference is generally nonstationary behaviour of the IF singularities, as we will show in the next Section. 
Thus, the vector field singularities are pure property of instantaneous field configuration. From the Eq. \eqref{GrindEQ__32_}, one can see that in the vicinity of IF singularities

\begin{equation} \label{GrindEQ__33_} 
\begin{array}{l} {\oint \textbf{E}_{\rho }  \textbf{n}dl=c \oint \textbf{B}_{\varphi }  d\textbf{l}} \\\\ {c \oint \textbf{B}_{\rho }  \textbf{n}dl=-\oint \textbf{E}_{\varphi }  d\textbf{l}}, \end{array} 
\end{equation} 
  where \textbf{n} is the unit vector normal  to the infinitesimal element of the loop \textit{dl}. The line integral of fields is calculated over the closed equiphase loop coinciding with a circle of equal intensity in the beam transverse section. 
These integrals are equal to zero for any subgroups except for \textit{l}$\sigma$=$\mathrm{-}$1 because $\textbf{g}_\rho$ and $\textbf{g}_\varphi$ components of vector function
\begin{equation} \label{GrindEQ__34_} 
\textbf{g}(\varphi )=\cos \left[(l+\sigma )\varphi +\varphi _{0} \right]\textbf{e}_{\rho } -\sigma \sin \left[(l+\sigma )\varphi +\varphi _{0} \right]\textbf{e}_{\varphi }  
\end{equation} 
change sign 2(\textit{l}+$\sigma$) times along  the closed loop. Nominally, such field configuration is similar to two-dimensional vector fields of \textit{N}=2{\textbar}\textit{l}+$\sigma${\textbar}  charge``sources'' (or currents) symmetrically localized in the vicinity of the axis $\rho=0$ for the\textit{ l}$\sigma$$<$0  and at infinity $\rho\rightarrow\infty$ for \textit{ l}$\sigma$$>$0. The parity of these sources leads to cancellation  of the loop integrals Eq.(\ref{GrindEQ__33_}) in the vicinity of the IF singularity.

The exceptional subgroup \textit{ l}$\sigma$$=$-1 is unique as the vector field \textbf{g} retains  fixed orientation along the loop. This corresponds to a spirally oriented vector field \textbf{g} with two extreme cases $\varphi=0$ and $\varphi=\pi/2$ of pure radial and azimuthal orientations when the function \eqref{GrindEQ__34_} is independent of the coordinate $\varphi$

\begin{equation} \label{GrindEQ__35_} 
\textbf{g}=\textbf{e}_\rho,\quad\mbox{when } {\varphi_0}=0,  
\end{equation}
and
\begin{equation} \label{GrindEQ__35_a} 
\textbf{g}=\textbf{e}_\varphi,\quad\mbox{when } \varphi_0=\pi/2. 
\end{equation}

\begin{figure}[htbp]
\centering
\includegraphics[width=8.0cm]{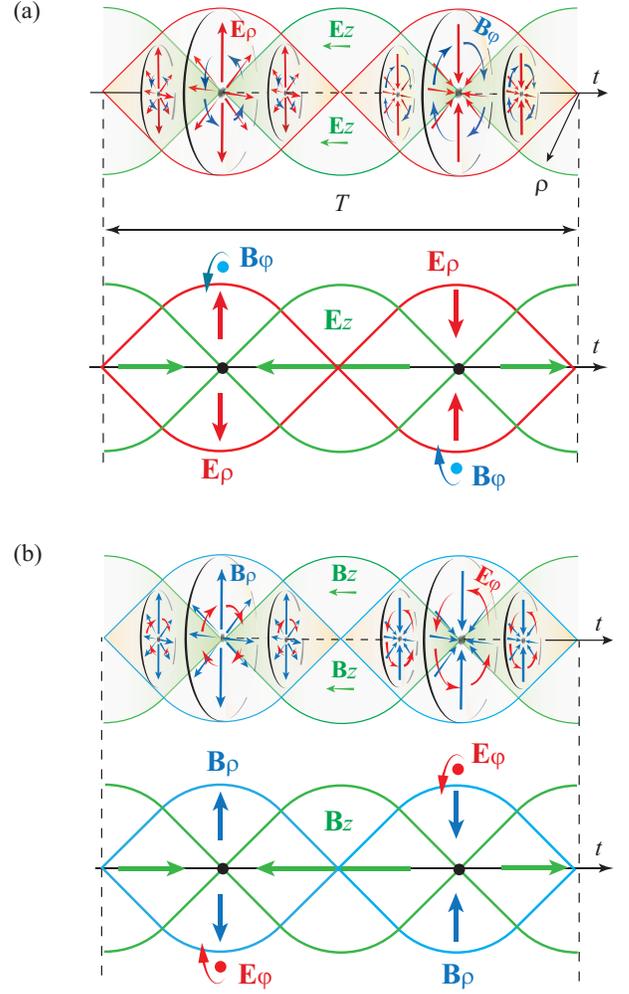}
\caption{\label{Fig4} \@  Field distributions over the wave period of the radially (a) and azimuthally (b) polarized vector beams. Instantaneous field singularities 
manifest themselves as virtual sources and sinks of a vector fields in points of zero field amplitude. 
Each source/sink of the transverse field is, at the same time, a sink/source of the longitudinal component of the field.}
\end{figure}

Electric fields associated with these modulating functions correspond to axially symmetric, radially and azimuthally polarized beams, respectively, known in optics as TM and TE beams (see [27] and references therein). 
\noindent The electromagnetic field of the TM and TE beams are

\begin{equation} \label{GrindEQ__37_} 
\begin{array}{c} {\textbf{E}^{TM}_{\bot } =E_{0} Re\left\{f(R_{t} ,\xi )Ge^{i(kz-\omega t)} \right\}\textbf{e}_{\rho } } \\ {\textbf{B}^{TM}_{\bot } =B_{0} Re\left\{f(R_{t}^{} ,\xi )Ge^{i(kz-\omega t)} \right\}\textbf{e}_{\varphi } } ,\end{array} 
\end{equation} 
and
\begin{equation} \label{GrindEQ__38_} 
\begin{array}{c} {\textbf{E}^{TE}_{\bot } =E_{0} Re\left\{f(R_{t} ,\xi )Ge^{i(kz-\omega t)} \right\}\textbf{e}_{\varphi } } \\ {\textbf{B}^{TE}_{\bot } =-B_{0} Re\left\{f(R_{t} ,\xi )Ge^{i(kz-\omega t)} \right\}\textbf{e}_{\rho } } .\end{array} 
\end{equation}

\noindent As it is clearly seen, the field pattern has uncompensated instantaneous radial component of electric (for the TM)  and magnetic (for the TE) field on any closed circular loop in the beam transverse section.  Such pattern corresponds to intriguing field configurations (see Fig. 4) as the integrals in Eq. 33 are not equal to zero on the loop.  Formally, according to the integral form of Maxwell's equations, this means existence of uncompensated charges (monopoles) and their currents in the vicinity of the IF singularity. To demonstrate this, let us consider the fields \eqref{GrindEQ__37_} and \eqref{GrindEQ__38_} at the plane\textit{ z}~=~0 near the point \textit{R${}_{t}$${}_{~}$}=~0. Taking into account that \textit{f~}=~\textit{R${}_{t}$${}_{~}$}${}^{ }$and using the  first-order Taylor series approximation of the field functions near the axis $\rho = 0$, the both \eqref{GrindEQ__37_} and \eqref{GrindEQ__38_} fields approximately are:

\begin{equation} \label{GrindEQ__39_} 
\begin{array}{c} {\textbf{E}^{TM}_{\bot}\approx E_{0} R_{t} \cos \omega t\; \textbf{e}_{\rho } } \\\\ {\textbf{B}^{TM}_{\bot}\approx B_{0} R_{t} \cos \omega t\; \textbf{e}_{\varphi } }, \end{array} 
\end{equation} 
and
\begin{equation} \label{GrindEQ__40_} 
\begin{array}{c} {\textbf{E}^{TE}_{\bot}\approx E_{0} R_{t} \cos \omega t\; \textbf{e}_{\varphi } } \\\\ {\textbf{B}^{TE}_{\bot}\approx -B_{0} R_{t} \cos \omega t\; \textbf{e}_{\rho } }. \end{array} 
\end{equation}

\noindent Consider a closed circular loop of equal field amplitudes in the \textit{z=~0 }plane with the centre at a point $\rho= 0$ and a radius \textit{R} as shown in Fig.5.  Let $\textbf{l}=\textbf{e}_\varphi$ be a unit vector alone the loop and  $\textbf{n}=\textbf{e}_\rho$ be a unit vector normal to the infinitesimal element $\textit{dl}$ of the loop. Then  the electric and magnetic field fluxes in radial direction through the loop, for   \textit{R}$\mathrm{\to}$0,  are 
\begin{equation} \label{GrindEQ__41_} 
\begin{array}{c} {\lim _{\Delta S\to 0} \frac{\oint \textbf{E}^{TM}_{\bot} \textbf{n} dl}{\Delta S} \approx (2E_{0} /w_{0} )\cos \omega t} \\\\ {\lim _{\Delta S\to 0} \frac{\oint \textbf{B}^{TE}_{\bot} \textbf{n} dl}{\Delta S} \approx -\; (2B_{0} /w_{0} )\cos \omega t}  ,\end{array} 
\end{equation}
 where we used  the fact that  the   loop length   $\mathrm{\oint}\textit{dl}=2{\pi}R$  and area, $\Delta S=\pi\textit{R}^2$. 
 On the other hand,  as the electric (magnetic) field flows through the closed loop, the dual to it magnetic (electric) field circulates around the loop:

\begin{equation} \label{GrindEQ__42_} 
\begin{array}{c} {\lim _{\Delta S\to 0} \frac{\oint \textbf{B}^{TM}_{\bot}  d\textbf{l}}{\Delta S} \approx (2B_{0} /w_{0} )\cos \omega t} \\\\ {\lim _{\Delta S\to 0} \frac{\oint \textbf{E}^{TE}_{\bot}  d\textbf{l}}{\Delta S} \approx (2E_{0} /w_{0} )\cos \omega t} . \end{array} 
\end{equation}
The expressions \eqref{GrindEQ__41_} can be  converted into the field fluxes through the closed infinitesimal cylindrical surface $\Omega$ with the length $\textit{c} \Delta\textit{t}$ along the \textit{z} direction and the volume $\Delta V=\textit{c}\Delta\textit{t} \,\Delta\textit{S}$ in the time interval $\textit{t}+\Delta \textit{t}$, as: 
 
\begin{equation} \label{GrindEQ__43_} 
\begin{array}{c} \oint \textbf{E}^{TM}_{\bot} \textbf{n} d\Omega =c\int _{t}^{t+\Delta t}\oint \textbf{E}^{TM}_{\bot} \textbf{n} dl  dt \\\\ \approx 2\pi cR^{2} (E_{0} / w_{0} )\int _{t}^{t+\Delta t}\cos \omega tdt \\\\ = 2\pi cR^{2} (E_{0} /\omega w_{0} )[\sin(\omega (t+\Delta t)) - \sin(\omega t)] \\\\ \oint \textbf{B}^{TE}_{\bot} \textbf{n} d\Omega =c\int _{t}^{t+\Delta t}\oint \textbf{B}^{TE}_{\bot}\textbf{n} dl  dt \\\\ \approx -2\pi cR^{2} (B_{0} /w_{0} )\int _{t}^{t+\Delta t}\cos \omega tdt \\\\ = 2\pi cR^{2} (B_{0} /\omega w_{0} )[\sin(\omega t)-\sin(\omega (t+\Delta t))]\end{array} 
\end{equation} 

We may assume that $\Delta$\textit{t} is very small such that the time variation of fields is nearly uniform over the cylindrical area. Then the electric and magnetic fluxes through the closed cylindrical surface divided by the volume $\Delta V$ are given by

\begin{equation} \label{GrindEQ__44_} 
\begin{array}{c} {\lim _{\Delta {\rm V}\to 0} \frac{\oint \textbf{E}^{TM}_{\bot}\textbf{n}\, d\Omega }{\Delta {\rm V}} \approx (2E_{0} \omega {\Delta t}/w_{0} )\cos \omega t} \\\\ {\lim _{\Delta {\rm V}\to 0} \frac{\oint \textbf{B}^{TE}_{\bot}\textbf{n}\, d\Omega }{\Delta {\rm V}} \approx -(2B_{0} \omega {\Delta t}/w_{0} )\cos \omega t} \end{array} 
\end{equation} 

\noindent Formally, according to Eqs.(42) and (44), the IF singularities Eq.\eqref{GrindEQ__35_} and Eq.\eqref{GrindEQ__35_a}, are source of non-compensated fluxes and circulations of the transverse  fields through the side area of the cylindrical surface. This is why we named corresponding  IF singularities as topological monopoles and currents. The field oscillations periodically change signs of the fluxes Eq.(44) and circulations Eq.(42) in any fixed plane section \textit{z}~=~const leading to zero average values over the wave period \textit{T}, as it shown in the Fig.4. These fluxes and circulations remain finite with the same signs between two zero points on a half period \textit{T}/2 of oscillations (or on a half wavelength distance along z-axis).

Note, that this  flux anomaly appearing in the transverse components of the wave field is removed by including the longitudinal field components as their fluxes through the closed surface (here through bases of the cylindrical surface) exactly compensate the  fluxes of the transverse fields. Although in the zero-order paraxial approximation the longitudinal \textit{z}-components of electric and magnetic fields are ignored, they appear in the next, first-order paraxial approximation as $\omega \; F_{z} =i\; c\nabla _{\bot } F_{\bot } $[27]. As an example, let us consider the longitudinal field flux through the plane z=0 in the radially polarized beam of Eq. 39. In such a beam the electric field with the complex function  $F_{z} =2icE_{0} /\omega w_{0} $ represents the amplitude of  longitudinal field, and is defined by Eqs.\eqref{GrindEQ__37_} and \eqref{GrindEQ__39_} as:

\begin{equation} \label{GrindEQ__45_} 
E_{z} =2E_{0} \sin (\omega t)/kw_{0}  
\end{equation}

The longitudinal field Eq. (45) is out of phase with respect to  the transverse components Eq. (39) and its  orientation is 
shown in the Fig. 4. The resulting  electric field flux of the longitudinal field through the cylindrical bases $\Delta\textit{S}$ is

\begin{equation} \label{GrindEQ__46_} 
\begin{array}{c} \oint \textbf{E}^{TM}_{\textit{z}} \textbf{n} d\Omega=\\\\ = -2\pi cR^{2} (E_{0} /\omega w_{0} )[\sin(\omega (t+\Delta t)) -\sin(\omega t)],\end{array}  
\end{equation} 
and exactly cancels  the flux  of the transverse components of the electromagnetic field depicted by  Eq. (43) . 

Figure 5 schematically shows fluxes and circulations of electric \textbf{E} and magnetic \textbf{B} fields on the cylindrical surface around the beams axes with the length of side equal
 to a fraction of wavelength $\Delta\lambda$.

\begin{figure}[htbp]
\centering
\includegraphics[width=8.0cm]{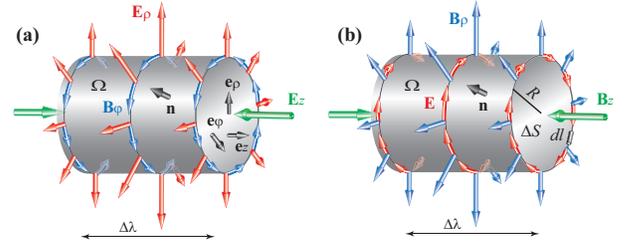}
\caption{\label{Fig5} \@  Fluxes and circulations of the electric $\textbf{E}_{\bot}$ and magnetic $\textbf{B}_{\bot}$ field at the fraction of wavelength $\Delta\lambda$ in radially (a) and azimuthally (b) polarized vector beams. In the radially (azimuthally) polarized beam, the transverse field $\textbf{E}_{\rho}$ ($\textbf{B}_{\rho}$) has nonzero flux through the side area of a closed cylindrical surface $\Omega$, while the field $\textbf{B}_{\varphi}$ ($\textbf{E}_{\varphi}$)  circulates around the closed loop 
on the surface. The flux of the longitudinal field $\textbf{E}_{\textit{z}}$ ($\textbf{B}_{\textit{z}}$) through the bases \textit{S} of radius \textit{R} compensates the transverse flux. The axis of the cylinder coincides with the beam axis. }
\end{figure}

The aforementioned  topological monopoles, as well as the other IF singularities of electromagnetic waves, are caused solely by the spatial configuration of the surrounding  transverse vector fields. 
However,  the full electromagnetic field of vector beams remains regular as all vector fields acquire physical values everywhere in the beams except the zero-field singular point.

\section{Polarization topology via moving \textit{IF} singularities}

 All considerations on  instantaneous vector singularities in  the previous section refer  to  vector beams with a uniform state of polarization  inhomogeneously distributed in space. In general, both orientation of the field vectors and ellipticity of the polarization can  vary from point to point. In three dimensions, when the state of elliptic polarization changes with position, points of circular polarization appear  along the so called C-lines, while the states of linear polarization with undetermined  handedness define   L-surfaces [25, 26, 52, 56].
 In the transverse cross-section, a L-surface is represented by the line. This line is swept out by at least one moving isolated singularity where \textit{instantaneous} electric and magnetic fields are zero, and their directions are therefore undetermined. These singularities  were first described by J. Nye and where originally named as ``wave disclinations'' [25]. Later, M. Dennis called them  ``relative singularities'' because their instantaneous position is phase-dependent [56] . 

The difference between the IF singularities in uniformly polarized beams, described above, and wave disclinations in beams with variant state of polarization is their dynamics. The IF singularities may have a stable spatial position and exist even in a wave with homogeneous state of polarization, while wave disclinations change their location in space over time sweeping out a L-surface. Although the instantaneous value of the electric (magnetic) field is equal to zero at points of wave disclinations, the field averaged over a period of wave oscillations has a nonzero value. This time-averaged field imprints a nonzero intensity on the L-surface on which the polarization becomes linear. As this is only a difference in the averaged picture, the wave disclinations are a particular case of IF singularities in our terminology. The morphology of a moving wave disclination belongs to the group $\textit{l}\sigma=\pm1$ of IF singularities and may be of source, spiral, sink, circulation or saddle morphology (see Fig. 3). The subgroup $\textit{l}\sigma=-1$  contains wave monopoles and currents of an instantaneous electromagnetic field in vicinity of singular points, similar as it did in the stationary case.   

The dynamic behaviour of the IF singularities affects  the time-averaged polarization states leading to polarization singularities - places related to C-lines and L-surfaces in beams with variation of  state of polarization. Because of their average nature, polarization singularities are not equal to the IF singularities and may even appear in points different from that of  IF singularities. The  dynamics of vector singularities can be simply illustrated through superposition of scalar and vector beams shown in previous Sections, as it leads to beams with typical polarization singularities.  More precisely, we combine \textit{m} vector modulating functions \eqref{GrindEQ__25_} and \eqref{GrindEQ__28_} with arbitral topological charges $\textit{l}_n$ and handedness $\sigma_n$

\begin{equation} \label{GrindEQ__47_} 
\textbf{U}=\sum _{n=1}^{m}A_{n} (\textbf{e}_{\rho } +i\sigma _{n} \; \textbf{e}_{\varphi } )f_n\; \exp (i(l_{n} +\sigma _{n} )\varphi ) ,     
\end{equation}
where 
 \textit{A${}_{n}$} is arbitrary constant. For simplicity, we ignore  initial phase in each function.

 The simplest superposition of the type \eqref{GrindEQ__47_} with nontrivial vector topology is sum of just two terms. The first term is a non-singular modulation function with the topological charge $\textit{l}_1=0$ and handedness $\sigma_1=-1,+1$. The second term is a function with topological charge \textit{l}${}_{2}$=$\mathrm{-}$1,+1 and handedness $\sigma_2=-1,+1$, which corresponds to a circularly polarised single charge vortex beam. The superposition of functions with the same handedness $\sigma _{1} \times \sigma _{2} =1$ leads to the case of homogeneously circularly polarized vortex beams considered in the previous Section. The difference is only the position of a singularity, which now is out of the optical axis. 
The other pair with opposite signs of handedness $\sigma _{1} \times \sigma _{2} =-1$ gives us vector beams with the following modulation functions

\begin{equation} \label{GrindEQ__48_} 
\begin{array}{l} \textbf{U}_{11} =A_1 (\textbf{e}_\rho +i\textbf{e}_\varphi)\exp (i\varphi )f_{1} \\\hspace{7mm}+ A_{2} (\textbf{e}_{\rho } -i\textbf{e}_{\varphi} )f_{2}  ,\\ 
 \\ 
 \textbf{U}_{12} =A_{1} (\textbf{e}_{\rho } -i\textbf{e}_{\varphi} )\exp (-i\varphi )f_{1} \\\hspace{7mm}+A_{2} (\textbf{e}_{\rho } +i\textbf{e}_{\varphi} )f_{2} \; \exp (2i\varphi ) , \end{array}  
\end{equation} 
and
\begin{equation} \label{GrindEQ__49_} 
\begin{array}{l} \textbf{U}_{21} =A_{1} (\textbf{e}_{\rho } -i\textbf{e}_{\varphi } )\exp (-i\varphi )f_{1} \\\hspace{7mm}+A_{2} (\textbf{e}_{\rho } +i\textbf{e}_{\varphi } )f_{2} , \\\\ \textbf{U}_{22} =A_{1} (\textbf{e}_{\rho } +i\textbf{e}_{\varphi } )\exp (i\varphi )f_{1} \\\hspace{7mm}+A_{2} (\textbf{e}_{\rho } -i\textbf{e}_{\varphi } )f_{2} \; \exp (-2i\varphi ) . \end{array}  
\end{equation}

The pairs (\textbf{U}${}_{11}$, \textbf{U}${}_{12}$) and (\textbf{U}${}_{21}$,\textbf{U}${}_{22}$) represent functions with opposite chiralities and can be transformed into each other via mirror reflection: $\sigma \to -\sigma ;\; l\to -l$. Hence, it is sufficiently to consider only the field  \eqref{GrindEQ__48_}. Vector beams represented by this superposition are named Poincar\'{e} beams because their polarization states cover entire surface of the Poincar\'{e} sphere [71-75]. If we choose the functions $f_{n} =R_{t}^{|l_{n} |} $, then the electric fields of both beams are
 
\begin{equation} \label{GrindEQ__50_} 
\begin{array}{l} \textbf{E}_1 =\text{Re}\;[E_{01} (\textbf{e}_{\rho } +i\textbf{e}_{\varphi } )\exp (i\varphi )\\\hspace{6mm}+E_{02} R_{t} (\textbf{e}_{\rho } -i\textbf{e}_{\varphi } )]G(\rho ,z)e^{i(kz-\omega t)} \\\\ \textbf{E}_2 =\text{Re}\;[E_{01} (\textbf{e}_{\rho } -i\textbf{e}_{\varphi } )\exp (-i\varphi )\\\hspace{6mm}+E_{02} R_{t} (\textbf{e}_{\rho } +i\textbf{e}_{\varphi } )\exp (2i\varphi )]G(\rho ,z)e^{i(kz-\omega t)}  ,\end{array} 
\end{equation} 
where \textit{E${}_{0n}$} are arbitrary amplitude constants. The Bessel-Gauss type of Poincar\'{e} beams can be obtained in a similar way by taking $\textit{f}=\textit{J}_{|l_n|}(KR_t)\textit{Z}(\xi)$.

The electric field evolution of  the beams given by Eqs. \eqref{GrindEQ__50_} is depicted  in Fig. 6 (a, b). At any moment of time, the isolated point of the IF singularity lies on a L-line in the transverse plane of the beams. This point rotates completing a full circle along the L-line during a half period of field's oscillations. At the same time  the IF singularity experiences a periodic transformation of  its structure (see the bottom panels in Fig. 6) similar to that of stationary IF singularities in circularly polarized vortex beams (see Fig.2).
\begin{figure}[htbp]
\centering
\includegraphics[width=8cm]{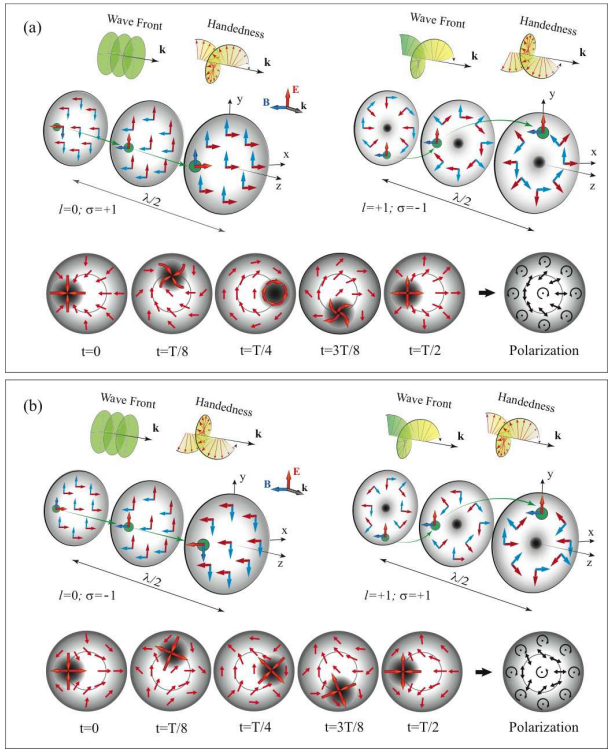}
\caption{\label{Fig6} \@  Field distribution and its dynamics in superposition [Eq.(48)] of circularly polarized Gaussian (left graph) and vortex beams (right graph) when a) $l_1$=0; $\sigma_1$=1; $l_2$=1;  $\sigma_2$=-1, and b) $l_1$=0;  $\sigma_1$  =-1; $l_2$=1;  $\sigma_2$  =+1. The evolution of a fixed arbitrary point on the wave surface is shown with a green circle. 
The bottom row in (a) and (b) panels depicts the  snapshots of electric component  of the field  taken in few time instants,  as well as the resulting  polarization structure of the beam. }
\end{figure}

  In three dimensions the distribution of the instantaneous field of Poincar\'{e} beams \eqref{GrindEQ__50_} with a single topological charge contains a single line of IF singularity where the field is discontinuous. This line sweeps out the L-surface with the field oscillation as the IF singularity rotates with double wave frequency around the beam axis while the axis becomes a C-line with the circular state of polarization. As topological charge of Poincar\'{e} beams \eqref{GrindEQ__50_} is uncompensated, these beams carry orbital angular momentum and the whole polarization pattern rotates in propagation. 

It is worth to note that unlike L-surfaces, C-lines are free from physical discontinuities of electromagnetic fields and have no IF singularities. The C-lines result from dynamic behaviour of the IF singularities on the L-surface. The topology of the polarization patterns in the vicinity of the C-point (transverse section of the C line) is related to average field orientations rather than to the field discontinuous. Those patterns are typically named as ``lemon'', ``star'' and ``monstar'' [56, 77]. Classification of C-points constitutes an extended topic discussed extensively in the literature (see for instance [49-56, 77-79] and review [9]. As they do not  possess the IF singularities, which are  our main interest here, we only remark that Poincar\'{e} beams given by Eqs.(50) and demonstrated in Fig.6 (a),  and (b), contain a ``star'' and a ``lemon'' C-points, respectively.

In closing of this Section let us note that we have chosen one of the simplest possible superposition of modulating functions \eqref{GrindEQ__28_} to clarify the instantaneous field singularities on polarization singularities. The selected superposition of the scalar function (corresponding to circularly polarized Gaussian beam) and the vector function (corresponding to circularly polarized vortex beam) is not the only one covering all states of polarization and leading to polarization singularities. The other simplest type of Poincar\'{e} beams can be constructed by the following superposition [80]
\begin{equation}
 \textbf{U}=A_{1} \textbf{e}_{x} +A_{2} R_{t} \textbf{e}_{y} \exp (\pm i\varphi ).
\end{equation} 
Each   of the constituent  functions is a scalar by itself as it represents a single   Cartesian component, but their superposition is a fully vectorial function (see Eq.(9)).

\section{Conclusions}
\noindent We presented here  systematic description of topological singularities in both electric and magnetic field of the electromagnetic waves, in a paraxial approximation. In particular,  we emphasized the basic physical principles underlying the role of instantaneous singularities. 
It appears that singularities essentially define a spatial topology of the surrounding wave field but not the evolution of wave packet in space and hence these two aspects can be analyzed separately. The envelope function generates scalar ``wave space" for the existence of phase, polarization or instantaneous vector field singularities defining the finite energy flux of electromagnetic field. The representation of the singularities by a single abstract modulating function greatly simplifies their formal description and, consequently, understanding of their fundamental nature.

The undefined direction of electromagnetic field vectors at points of singularities leads to paradoxical effect -- the instantaneous spatial distributions of the transverse field components around the singularities appear as if they were originating from 2D ``virtual'' electric or magnetic sources. These apparent topological monopoles are result of the particular spatial configuration of the transverse components of the electromagnetic field. 

\noindent
\section{Acknowledgment}
 The authors acknowledge the support from the Australian Research Council and the Qatar National Research Fund (grant NPRP8-246-1-060). We also thank the Anonymous  Referee  who  helped  us to significantly improve our work. 
 

\end{document}